\documentclass[sigconf]{acmart}

\usepackage{amsmath,amsfonts}
\usepackage{algorithmic}
\usepackage{graphicx}
\usepackage{textcomp}
\usepackage{xcolor}
\def\BibTeX{{\rm B\kern-.05em{\sc i\kern-.025em b}\kern-.08em
    T\kern-.1667em\lower.7ex\hbox{E}\kern-.125emX}}
\usepackage[utf8]{inputenc}
\usepackage{booktabs} 
\usepackage{graphicx}
\usepackage{subfigure}
\usepackage{units}
\usepackage{blindtext}
\usepackage{xcolor,xspace}
\usepackage{amsmath}
\usepackage{paralist}
\usepackage{mdframed}
\usepackage{pict2e}
\usepackage{mdframed}
\usepackage{listings}
\usepackage{color}
\usepackage{threeparttable}
\usepackage{array}
\usepackage{booktabs}
\usepackage{pifont}
\usepackage{balance}
\usepackage{url}

\usepackage[ruled]{algorithm2e}
\usepackage{algorithmic}
\LinesNumbered

\usepackage{url}

\usepackage[
    n,
    advantage,
    operators,
    sets,
    adversary,
    landau,
    probability,
    notions,
    logic,
    ff,
    mm,
    primitives,
    events,
    complexity,
    asymptotics,
    keys]{cryptocode}

\newcommand{\styleEntity}[1]  {\ensuremath{\mathsf{\texttt{#1}}}\xspace}

\newcommand{\BT}{\ensuremath{\styleEntity{BT}}\xspace}
\newcommand{\BTs}{\ensuremath{\styleEntity{BTs}}\xspace}

\newcommand{\FV}{\ensuremath{\styleEntity{FV}}\xspace}
\newcommand{\FVs}{\ensuremath{\styleEntity{FVs}}\xspace}
\newcommand{\FE}{\ensuremath{\styleEntity{FE}}\xspace}
\newcommand{\FEs}{\ensuremath{\styleEntity{FEs}}\xspace}
\newcommand{\HD}{\ensuremath{\styleEntity{HD}}\xspace}

\newcommand{\sadv}{{\ensuremath{\sf{\mathcal Adv}}}\xspace}
\newcommand{\vrf}{{\ensuremath{\sf{\mathcal Vrf}}}\xspace}
\newcommand{\chal}{{\ensuremath{\sf{\mathcal Chal}}}\xspace}

\newcommand{\rot}{{\ensuremath{\sf{\mathcal RoT}}}\xspace}
\newcommand{\problem}{{\ensuremath{\sf{RTI}}}\xspace}
\newcommand{\dev}{{\ensuremath{\sf{\mathcal Dev}}}\xspace}

\newcommand{\devA}{{Dev-A}\xspace}
\newcommand{\devB}{{Dev-B}\xspace}
\newcommand{\devC}{{Dev-C}\xspace}

\newcommand{\prv}{{$\rot_P$}\xspace}
\newcommand{\rotA}{{$\rot_A$}\xspace}
\newcommand{\rotB}{{$\rot_B$}\xspace}
\newcommand{\rotC}{{$\rot_C$}\xspace}

\newcommand{\devAdv}{{\ensuremath{\sf{\mathcal \dev^*}}}\xspace}
\newcommand{\rotAdv}{{\ensuremath{\sf{\mathcal \rot^*}}}\xspace}
\newcommand{\rpki}{\text{$pk_i$}\xspace}
\newcommand{\rski}{\text{$sk_i$}\xspace}

\newcommand{\rpkiA}{\text{\rotA-$pk(i)$}\xspace}
\newcommand{\rskiA}{\text{\rotA-$sk(i)$}\xspace}

\newcommand{\protocol}{{\ensuremath{\sf{RTI}}}\xspace}

\newtheorem{Definition}   {\textbf{Definition}}

\newtheorem{Theorem}   {\textbf{Theorem}}

\newtheorem{Proof}   {\textbf{Proof} (Sketch)}

\AtBeginDocument{%
  \providecommand\BibTeX{{%
    \normalfont B\kern-0.5em{\scshape i\kern-0.25em b}\kern-0.8em\TeX}}}

\setcopyright{acmcopyright}
\copyrightyear{2018}
\acmYear{2018}
\acmDOI{10.1145/1122445.1122456}

\acmConference[Woodstock '18]{Woodstock '18: ACM Symposium on Neural
  Gaze Detection}{June 03--05, 2018}{Woodstock, NY}
\acmBooktitle{Woodstock '18: ACM Symposium on Neural Gaze Detection,
  June 03--05, 2018, Woodstock, NY}
\acmPrice{15.00}
\acmISBN{978-1-4503-XXXX-X/18/06}

\begin{document}

\title{On the Root of Trust Identification Problem}

\author{Ivan De Oliveira Nunes}
\affiliation{%
  \institution{UC Irvine}}

\author{Xuhua Ding}
\affiliation{%
  \institution{Singapore Management University}}
\author{Gene Tsudik}
\affiliation{%
  \institution{UC Irvine}}

\begin{abstract}
Trusted Execution Environments (TEEs) are becoming ubiquitous and are currently used in many
security applications: from personal IoT gadgets to banking and databases. Prominent examples of 
such architectures are Intel SGX, ARM TrustZone, and Trusted Platform Modules (TPMs).
A typical TEE relies on a dynamic Root of Trust (RoT) to provide security services such 
as code/data confidentiality and integrity, isolated secure software execution, remote attestation, and sensor auditing.
Despite their usefulness, there is currently no secure means to determine whether a given security service or task is being 
performed by the particular RoT within a specific physical device. We refer to this as the Root of Trust Identification 
(\problem) problem and discuss how it inhibits security for applications such as sensing and actuation.

We formalize the \problem problem and argue that security of \problem protocols is especially challenging due to local 
adversaries, cuckoo adversaries, and the combination thereof. To cope with this problem we propose a simple and
effective protocol based on biometrics. Unlike biometric-based user authentication, our approach is not concerned with 
verifying user identity, and requires neither pre-enrollment nor persistent storage for biometric templates.
Instead, it takes advantage of the difficulty of cloning a biometric in real-time to securely identify the RoT of a given 
physical device, by using the biometric as a challenge. Security of the proposed protocol is analyzed in the 
combined Local and Cuckoo adversarial model. Also, a prototype implementation is used to demonstrate the protocol's 
feasibility and practicality. We further propose a Proxy \problem protocol, wherein a previously identified RoT assists a 
remote verifier in identifying new RoTs.
\end{abstract}




\maketitle


\section{Introduction}\label{sec:intro}
In recent years, there has been a growing demand, from both industrial and research communities, for
Trusted Execution Environments (TEEs) to aid security-critical applications.
While TEEs vary widely in terms of architecture, implementation, and functionality, they 
provide (at least in the idealized model) an isolated execution space offering both code and 
data integrity, without relying on any assumptions about applications or operating systems. 
We refer to these functionalities as TEE services. Security of TEE services (among other trusted services) rely on dynamic Roots of Trust (RoTs) to prove their integrity.
RoTs consist of minimal trusted components (e.g., trusted hardware as in TPM and Intel SGX, or trusted software as in hypervisors) used to bootstrap and dynamically verify trust in the system as a whole.

Despite the popularity of such services, it is somewhat surprising that there are no ``off-the-shelf'' means to securely bind 
a given RoT to the specific physical device housing this RoT.
In particular, it is easy to verify that 
a service is indeed performed by \textit{\textbf{some}} RoT. However, it remains a challenge to 
determine if the service is performed by \textit{the} RoT residing inside a specific physical device. We refer to this problem as Root of Trust Identification (\problem).

To further illustrate and motivate \problem, consider the following sensor auditing scenario highlighted in~\cite{ditio}.
 A device (e.g., a smartphone) keeps a TEE-enabled secure log of its audio and video (camera and microphone)
activity in order to allow after-the-fact auditing. For example, the host of a confidential meeting uses her trusted verifier device to verify 
that microphones and cameras of attendees' smartphones remain turned off.
The technique proposed in~\cite{ditio} consists of using each attendee device's TEE to assure (e.g., via remote attestation) the verifier of the integrity
of sensor usage logs on that device. We argue that -- even with TEE-based integrity assurance -- the attendee can still use his device's 
microphone/camera and fool the verifier by supplying logs from a remote accomplice device (also equipped with a TEE of same type) 
that indeed turns off the sensor during the meeting. The response appears to be valid and there is no means for the host to 
differentiate between replies from the accomplice device and the one presently held by the malicious attendee. Using a dedicated physical channel (e.g., a cable) between the 
verifier and the attendee's device does not solve the problem as the device may use another channel to communicate with its accomplice.  

Another scenario relevant to \problem occurs whenever some malware has been found on a device. A natural course of action
is to force one or more of: (i) re-set, (ii) update software, or (iii) erase the device. However, none of these is trivial since the same 
adversarial behavior can fool the user into believing that her device has been re-set/updated/erased, while in fact it is some
other device that has performed those actions.

\begin{table*}[!hbtp]
\centering
\Small
\caption{Notation summary}
\label{table:notation}
\begin{tabular}{|l|p{12.3cm}|}
\hline
Notation    					&  Description  								\\ \hline \hline
\devA, \devB, \devC, ...			&  Physical devices (e.g., smart-phone, laptop) A, B, C, ...	\\ &\\
\rotA, \rotB, \rotC, ...			&  \rot residing on physical devices A, B, C, ... \\ &\\
$\rpki, \rski \gets Gen(\text{\rotA})$		&  \rotA issues $i$-th session public-key \rpki, and corresponding secret key \rski. Anyone can verify that \rpki was generated by some \rot. However, \rotA signs \rpki using its master secret key in a group signature scheme, 
	thus one cannot tell whether \rpki was issued by \rotA or not.\\ &\\
$Pr[A|B]$					&  Probability of event $A$ given that event $B$ is true\\ &\\
$Pr[A|\neg B]$					&  Probability of event $A$ given that event $B$ is \underline{not} true\\ &\\
$l$						&  Security  parameter \\ &\\
$negl(.)$					&  a negligible function: $negl(l) \leq 1/2^l$\\ &\\
\hline \hline 
$\BT \gets \mathsf{\BT.Sample}_{}(U,\text{\devA})$ & Sampling of biometric template \BT from user $U$ performed by biometric sensor on physical device \devA. \\ & \\
$\HD \gets \mathsf{\FV_{GEN}}(\BT,\chal)$	& Generation of helper data \HD from biometric template $\BT$ and randomness \chal. \\ & \\
$\chal' \gets \mathsf{\FV_{OPEN}}(\BT',\HD)$	& Reconstruction of randomness $\chal'$ from helper data \HD  and biometric \BT'. \\ & \\
$\sigma \gets \mathsf{sign}_{sk}(M)$		& Signature result $\sigma$ of using $sk$ to sign message $M$. Implicitly we assume $\mathsf{sign}_{sk}$ to be a confidentiality preserving signature scheme, i.e., $M$ cannot be extracted from $\sigma$\\ & \\
$\mathsf{verify}_{pk}(\sigma) \equiv M$	& Verification of signature $\sigma$ on message $M$ for public key $pk$.  \\
\hline
\end{tabular}
\normalsize
\end{table*}

Due to lack of \problem solutions, attacks of this type are applicable to any TEE-dependent application which assumes that the 
TEE indeed resides on the device of interest.
More genrally, it applies to any service relying on physical presence of an RoT (either hardware-based or software-based) within a particular device.
A successful \problem verification can bind the public-key used by the 
RoT for remote attestation with its hosting device.  
However, the binding only has a long-lasting effect for RoTs using a device-specific persistent public key. 
For those privacy-friendly TEEs that use short-lived public keys certified with a group signature (such as Intel SGX), the binding is ephemeral. 
Hence, it is imperative to conduct \problem verification on a per-session basis for TEEs with privacy and unlinkability protection.



We observe that many TEE-enabled devices (e.g., laptops, tablets and smartphones) are equipped
with biometric sensors connected to the TEE via secure physical channels. Because biometric templates are sensitive and hard to revoke, this secure 
channel is used to secure the biometric template in case of a compromised application and/or operating system, while still allowing 
biometric authentication as shown in FIDO~\cite{FIDO}.
In this paper, we propose a low-burden user-aided approach to \problem. The basic idea is that the TEE vouches for the biometric template
securely obtained from the hard-wired sensor.  We {\bf do not use biometrics} to authenticate the user. Instead, a biometric is used as a challenge.
Security of our approach is based on the difficulty of cloning a human biometric (e.g., a fingerprint) in real-time during \problem verification. 
However, prior enrollment of a user's biometric is not required. We also do not use the same biometric in different sessions. 
Because it is used as a challenge, the only properties the biometric needs are: sufficient entropy and (real-time) unclonability, which 
biometrics used for user authentication are assumed to have.

In the rest of this paper, after formalizing \problem and describing the attack models, we construct a biometric-based \problem scheme.
We also prototype and evaluate our scheme using an \rot based on a trusted micro-hypervisor to demonstrate its practicality.
We consider \problem as a subtle and important issue, which has been mostly overlooked. 

\newcommand{\PA}{\ensuremath{\small{PA}}\xspace}

\begin{figure*}[ht]
\begin{mdframed}
\small
\begin{Definition}[\problem Protocol]\label{def:RTI}~\\
$\protocol(A,\text{\rpki})$ is a 2-party interactive protocol executed between \vrf and \prv.\\
\vrf selects a physical device \devA and \prv issues \rpki -- a session public-key.\\
The protocol outputs 1 if \vrf concludes that \rpki was issued by \rotA; or 0 otherwise.
\end{Definition}
\vspace{1mm}
\hrule
\vspace{1mm}
%
%
\begin{Definition}[\problem Completeness]\label{def:RTI_comp}~\\
\protocol is complete iff:\\
\begin{equation}
Pr[\protocol(\text{\devA}, \text{\rpki})|(\text{\rpki} \gets Gen(\text{\rotA}))] = 1-negl(l)
\end {equation}
where $l$ is  the security parameter and $negl$ is a negligible function.
\end{Definition}
\vspace{1mm}
\hrule
\vspace{1mm}
%
\vspace{1mm}
\begin{Definition}[\problem Security]\label{def:RTI_sec}~\\
\protocol is secure iff:\\
\begin {equation}
 Pr[\protocol(\text{\devA}, \text{\rpki})|\neg(\text{\rpki} \gets Gen(\text{\rotA}))] = negl(l) 
\end {equation}
where $l$ is  the security parameter and $negl$ is a negligible function.
\end{Definition}
\end{mdframed}
\end{figure*}
\section{\problem Protocols}\label{sec:RTI}
In this section we define \problem protocols and the adversarial model.
Our notations are summarized in Table~\ref{table:notation}.
As noted in Section~\ref{sec:intro}, some types of TEEs use a device-specific persistent public key while others use one-time public key with group signature based certification.
Without loss of generality, our treatment in this section focuses on the latter type since it subsumes the former.
%
\subsection{Definitions}
Suppose \devA is a physical device (e.g., smartphone, personal computer, server) equipped with an RoT denoted by \rotA.
Let:
\begin{equation}
\rpki, \rski,\sigma_i \gets Gen(\text{\rotA})
\end{equation}
denote the process whereby \rotA generates the $i$-th asymmetric key pair $(\rpki,\rski)$ and a group signature $\sigma_i$ upon $\rpki$. 
Although $\sigma_i$ can be verified cryptographically, it does not prove that $\rpki$ is for \devA, because the signature does not enclose any physically identifiable property of \devA.


An \problem protocol is the interactions between a verifier (\vrf) and a prover RoT (\prv) which issues \rpki and is alleged to reside on \devA.
Both parties are trusted and cooperate such that \vrf can decide if \prv resides in \devA, 
i.e., whether \prv $\equiv$ \rotA. Interestingly, not even \prv itself knows its own residency. This goal is deceptively simple 
and, as we discuss in the remainder of this paper, is hard to achieve even though both parties involved in the 
protocol are trusted.
 
At the end of the \problem protocol, \vrf learns \rpki which is a public key used by \prv. \vrf's assertion on \prv $\equiv$ \rotA also implies that \rpki is indeed issued by \rotA.
Completeness and security of \problem are defined in terms of \vrf's ability to make a positive conclusion
if and only if $\rpki \gets Gen(\text{\rotA})$, with overwhelming probability.
We specify a generic $\protocol$ protocol in Definition~\ref{def:RTI}.
Completeness and security of \problem protocols are stated in Definitions~\ref{def:RTI_comp} and~\ref{def:RTI_sec}, respectively.

Definition~\ref{def:RTI_comp} states that a complete \problem protocol against \rotA, always outputs
`1'  if the public-key \rpki given as input to the protocol is indeed generated by \rotA.  Definition~\ref{def:RTI_sec} 
states that a secure \problem protocol against \rotA, always output `0',  if \rpki given as 
input to the protocol is not issued by \rotA. Note that by Definition~\ref{def:RTI}, \prv is defined as 
the \rot that issues \rpki, thus the following equivalence:
%
\begin{equation}
[\text{\rpki} \gets Gen(\text{\rotA})] \leftrightarrow [\text{\prv} \equiv \text{\rotA}].
\end{equation}
%
We now present several possible attacks on \problem protocols to illustrate some subtleties in addressing
the \problem problem.

\subsection{Attack Vectors}\label{sec:vectors}
\begin{figure*}[!htbp]
\centering
  \subfigure[Expected setting in benign \problem protocol execution.]
  {\includegraphics[width=3.6cm,height=1.0in]{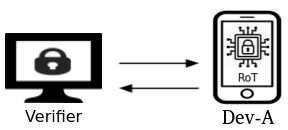}\label{fig:1a}} \hspace{1cm}
  \subfigure[An evil-twin \sadv uses \devAdv to hijack the communication and play the role of \devA.]
  {\includegraphics[width=5.8cm,height=1.2in]{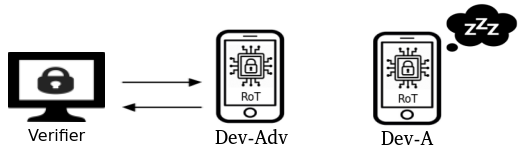}\label{fig:1b}} \hspace{1cm}
  \subfigure[A cuckoo \sadv uses malware on \devA to relay \vrf messages to/from accomplice \devAdv.]
  {\includegraphics[width=5.8cm,height=1.0in]{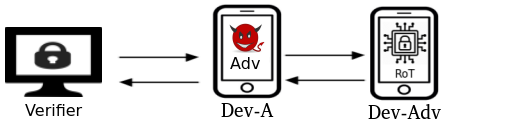}\label{fig:1c}}
\caption{Possible scenarios during \problem protocol execution}\label{fig:settings}
\end{figure*}
In this section, we discuss several attack scenarios and argue that addressing \problem is challenging.
We start by describing a na\"ive approach to solving \problem and show how it can be attacked trivially.
We then gradually increase adversarial capabilities. 

\subsubsection{Na\"ive \problem Protocol}
As shown in \cite{ditio}, a natural way to solve \problem is to challenge whether \prv knows \rski, assuming that 
\vrf has the prior knowledge of  \rotA's ownership of $\rski$. The protocol supposes the scenario 
in Figure~\ref{fig:1a} and proceeds as follows (communication is assumed  to take place
over a wireless medium):
\begin{compactenum}
 \item \vrf requests \prv public key;
 \item \vrf receives \rpki and checks that it was issued by some legitimate \rot by verifying the group signature on \rpki;
 \item \vrf issues a random challenge $c$, encrypts $c$ under \rpki, and sends it to \prv;
 \item \prv issues signs $c$ using its private key;
 \item \vrf verifies the signature from \prv using \rpki. If valid, it concludes that \prv is \rotA and \rpki is indeed issued by \rotA;
\end{compactenum}
The problem is that the assumption in the na\"ive protocol barely holds in reality because it is infeasible for \vrf to have the prior knowledge of ownership of the key.
Hence, \vrf cannot distinguish between an interaction with \devA and 
some other \devAdv of the same class and equipped with the same \rot type.
In particular, an evil-twin adversary \sadv can easily convince \vrf that \rpki was issued by \rotA
while in fact \rpki is issued by \rotAdv. As illustrated in Figure~\ref{fig:1b}, \sadv performs as follows:
\begin{compactenum}
 \item \sadv intercepts \vrf request and forwards it to \devAdv; 
 \item \sadv replies to \vrf with $\text{\rpki} \gets Gen(\text{\rotAdv})$, issued by \rotAdv;
 \item \vrf believes that \rpki was generated by \rotA and completes the rest of the protocol with \rotAdv;
 \item \vrf incorrectly concludes that \rpki was issued by \rotA.
\end{compactenum}
\noindent\textit{\textbf{Remark}: Although \rotAdv is honest (i.e., not subverted by \sadv), 
it cannot tell that it is being (ab)used by \sadv to fool \vrf. From \rotAdv perspective, this interaction is 
indistinguishable from a legitimate execution of an \problem protocol between \vrf and itself.}

\subsubsection{Coping with Evil-twin Adversaries}
One way to cope with an evil-twin adversaries is for \vrf to require a physical channel that cannot 
be tampered with, or accessed, by nearby devices. For example, intercepting \vrf messages and replying in 
place of \devA is significantly harder when \vrf uses a wired channel (e.g., a USB cable) to communicate with \devA.
This would prevent \sadv from using \devAdv to interact with \vrf directly, since only \devA is directly 
connected to \vrf. In this case, an honest execution of the \problem protocol would proceed as 
above, except for the use of the wired channel.
However, even a wired channel is insufficient if we consider a \emph{cuckoo} adversary~\cite{parno2011bootstrapping}.
Such an adversary first installs malware on \devA.  This malware intercepts incoming messages 
destined for \rotA and forwards them to \devAdv. As illustrated by Figure~\ref{fig:1c}, the attack proceeds as follows:
\begin{compactenum}
 \item Malware on \devA forwards \vrf request (received on the direct channel) to \devAdv, which feeds to \rotAdv;
 \item \rotAdv replies to \vrf request. It issues a \rpki and plays its part in the 
 	challenge-response protocol with \vrf (inadvertently assuming the role of of \rotA);
 \item Response message from \rotAdv is relayed to \vrf by malware on \devA, via the direct channel.
 \item \vrf incorrectly concludes that \rpki is issued by \rotA.
\end{compactenum}
As in the evil-twin attack case, \rotAdv is an honest \rot. However, it cannot tell that it is used by \sadv to fool \vrf.

\subsubsection{Cuckoo Adversaries}
Cuckoo attacks show that defending against evil-twin adversaries is not enough when 
malware is in full control of \devA. Indeed, the threat of malware is the main reason for 
\devA to be equipped with an \rot. On the other hand, because network I/O interfaces 
typically go through untrusted components (i.e., drivers and OS),  malware presence 
makes a secure physical connection between \vrf and \devA insufficient for mitigating the \problem problem.
Capabilities of a cuckoo attacker are not restricted to the wired interface (e.g., USB). Any I/O device that does 
not communicate directly to the \rot must pass through an untrusted component and can be used for cuckoo attacks.

As a matter of fact, an \rot could even be used to verify the existence of a software direct secure path 
(e.g., implemented by a hypervisor) between itself and the I/O interface inside a given device.
Then, as a part of the \problem protocol, \rot would only reply to a challenge coming on that particular verified interface.
In the cuckoo attack, \rotAdv (which is an honest \rot) would refuse to reply to the challenge relayed by 
\sadv, because it is not received from the expected and verified wired I/O interface, since \devA and \devAdv are not directly connected.

Unfortunately, even this setting can be circumvented by a more potent cuckoo \sadv which uses an \emph{accomplice challenger} 
that connects to \devAdv via a channel expected by the \rot. Malicious software on \devA can forward \vrf messages to the 
\emph{accomplice challenger}. 
The \emph{accomplice challenger} then forwards to \devAdv the same messages sent by \vrf to \devA, over 
the expected I/O interface. Since the view of \rotAdv is indistinguishable from that of an honest execution of \problem, 
it produces a legitimate response that passes the verification. 

Although the channel expected by the \rot in our example is a wire/cable, 
this attack applies to any I/O interface. 
Assuming that the \emph{accomplice challenger} has I/O capabilities equivalent to those of \vrf, 
a challenge from \vrf can be replayed by the \emph{accomplice challenger} using the same type of channel.
Thus, we observe that \textbf{\emph{whenever the challenge is conveyed using a machine I/O interface, 
it can be replayed by another machine with the same I/O capabilities}}. This motivates our choice for a biometric-based 
\problem scheme. The key rationale is that, if a human user becomes a part of the I/O operation, this I/O operation 
cannot be easily replayed since it requires physical participation by the same person.

\subsection{\problem Adversarial Model}
Considering the attack scenarios of Section~\ref{sec:vectors}, we define a strong adversary 
\sadv that can compromise the entire software stack of \devA, \emph{excluding the software component of \rotA e.g., a trusted hypervisor loaded and verified by the hardware component of \rotA}.
As such, \sadv can compromise applications and the operating system. It can intercept, eavesdrop, discard or inject messages 
on the internal path between \devA's I/O interfaces and \rotA.

We assume that \sadv has the same capabilities (intercept, eavesdrop, discard or inject messages) on the network.
\sadv can sense physical surroundings of \devA and \vrf and record, retransmit, and replay any message, 
signal or action performed by \vrf or \devA actuators. In particular, \sadv can deploy its own sensors and actuators with 
I/O capabilities equivalent to those of \vrf and \devA, in the environment surrounding them.
This model accounts for \emph{evil-twin adversaries} as per Section~\ref{sec:vectors}.

\sadv can deploy an accomplice device \devAdv equipped with \rotAdv. The entire software state of \devAdv is also under \sadv control.
These devices might be located in a remote environment where \sadv deploys its own sensors and actuators with 
I/O capabilities equivalent to those of \vrf and \devA.
Malware on \devA (controlled by \sadv) might, for instance, intercept messages sent from \vrf to \rotA, relaying them to \devAdv. 
\rotAdv might inadvertently reply to malware on \devA which then forwards replies to \vrf on behalf of \rotA.
This model accounts for \emph{cuckoo adversaries}, as discussed in Section~\ref{sec:vectors}.

We consider hardware attacks to be out of scope of this paper. Specifically, \sadv\ cannot make hardware changes on \devA, 
any hardware-based \rot, or the physically built-in circuit linking a trusted I/O device and an \rot.
Protection against physical attacks is considered orthogonal and can be supported by 
tamper-resistant hardware techniques~\cite{ravi2004tamper}.

\subsection{Mitigating \problem via Presence Attestation}\label{sec:pres_att}
The \problem problem is quite similar to that of convincing a \emph{human user} that her own device has 
an {\em active} \rot. The latter is referred to as {\emph{Presence Attestation} (\PA)  
in~\cite{presence_att} which proposes several concrete schemes. In addition to convincing the human 
user that she is interacting with the \rot on her device, \PA schemes can be extended so that \vrf learns 
the \rot's public key. Therefore, they are one way to address \problem.
Unsurprisingly, \PA schemes also cope with evil-twin and cuckoo attacks.
We now overview three \PA schemes from~\cite{presence_att} and discuss their security from the 
\problem perspective.

\subsubsection{Location-based \PA}
The security premise of location-based \PA scheme is twofold: (i) \rot securely obtains genuine location of 
its hosting device, as reported by GPS; and (ii) given sufficient knowledge about \devA's location, the user 
can manually verify location reported by \rot, perhaps aided by visualization on a map.    
The essence of this approach is to use the geographic location as the challenge to \rot. However, besides 
well-known attacks on GPS signaling~\cite{GPS1,GPS2}, its main shortcoming is that it cannot differentiate 
\rotA from \rotAdv, which is sufficiently close to \devA so that they report the same readings.
Moreover, manual verification of a geographic location does not have high enough accuracy. 

\subsubsection{Scene-based \PA}
This scheme uses a (photo of a) scene randomly chosen by the human user as the 
challenge and requires \rot to report the challenge received over a secure camera interface. As in the location-based scheme, 
the human user verifies correctness of the \rot response.    
This scheme is vulnerable to the evil-twin attack where the adversary takes the picture of the same scene and asks \rotAdv to sign it. Its security is 
therefore dependent on the human user's ability to differentiate among photos taken by two different devices, which is 
obliviously not reliable. This scheme is also vulnerable to analog cuckoo attacks, whereby \sadv re-renders the scene to 
an accomplice display such that \devAdv can take a genuine photo of it. Given today's hardware technology, it is 
infeasible for a normal user to distinguish between a photo of a physical scene and a re-production thereof.  
In both location- and the scene-based schemes, the human user decides on correctness of \rotA's response. From the perspective 
of \problem, it takes an extra step for the user to notify \vrf about her conclusion.  

\subsubsection{Sight-based \PA}
Sight-based \PA scheme does not require any human input. Its security is based on the observation that any message 
reply in the line-of-sight channel incurs measurable time delay, because the attack includes analog operations which are 
comparatively time-consuming. In this scheme, \vrf and \devA run the standard challenge-response protocol using 
the line-of-sight channel whereby a display ``sends" messages to a camera. Using cryptographic means, \vrf checks 
integrity of the response. In addition, by measuring the time to complete the session, it verifies whether \rotA is at the 
other end of the light-of-sight channel. Note that this scheme requires \rotA to securely obtain the challenge from 
the camera and securely display the response to the display.  

Although it offers stronger security than location- and scene-based schemes, sight-based \PA is dependent on the 
current frame-per-second (fps) rate of commodity cameras on modern smartphones.
Moreover, sight-based \PA requires the two participating devices to be physically 
well positioned through multiple rounds in order to form a high-quality light-of-sight channel. 

\noindent\textbf{Summary.} Zhang et. al. \cite{presence_att} have shed light on challenges related to \rot and cuckoo attacks, and made 
attempts to tackle them. We believe that \problem is both harder and more general than the \PA problem, since \problem does
not assume that the average human user possesses sufficient knowledge  and expertise to discern ambient properties. Our 
biometric-based approach relies on the unclonability of human biometrics with high entropy, the same assumption propping 
up security in biometric authentication schemes.

\subsection{Mitigating RTI via Distance Bounding}
Distance bounding protocols~\cite{distbound1,distbound2,distbound3,distbound5} allow a verifier to determine whether its communication peer is within a certain distance (e.g., 30 cm). They are fundamentally different from a RTI protocol because establishing an acceptable distance does not always \emph{identify} the device. Using distance to solve RTI assumes that there is only single device in the range, which does not hold when the distance is large.   

There are also implementation issues using a distance-bounding protocol for RTI. Parno et. al \cite{parno2011bootstrapping} have remarked that it is not suited to deal with the cuckoo attack against TPM-based attestation given the slow speed of TPM. Although today's \rot has better performance, the time variance of signature generation remains too large for distance-bounding protocols which only tolerate time errors in several nanoseconds. Moreover, distance-bounding protocols would require all devices of \devA's class to be equipped with distance bounding hardware (ultra wide-band radios with high-precision clocks needed for accurate timing measurements) securely wired to the \rot ~\cite{distbound4}. 
This is currently not available in commodity devices. 

Recently, Dhar et. al. \cite{DPKC18} propose to use a trusted device (e.g., a smart USB device) as a proxy attached to the proving device so that a remote verifier detect the cuckoo attack during SGX attestation. Besides the hassle of using a trusted device, this approach relies on a strong assumption that the trusted device attached to an untrusted environment remain intact.   


%


\section{Building Blocks}\label{sec:preliminaries}
\subsection{Biometric Features \& Template Matching}
A Biometric Template (\BT) is composed of features uniquely identifying an individual. In a biometric application (e.g., user authentication) a 
reference \BT is usually sampled and stored as part of the enrollment procedure. During authentication, the feature extraction procedure 
is used to collect a real-time sample \BT' from the purported user. If the similarity score between \BT' and \BT exceeds a pre-defined threshold, 
they are considered as a matching pair. The method to evaluate the similarity score and the choice of the threshold depend on the particular 
biometric. A \BT corresponding to user $U$ is represented by a set:
\begin{equation}\label{bt_format}
\BT_U = \{p_1, ..., p_M\}\,,
\end{equation}
where $p_1, ..., p_M$ are data points (features) representing unique details of $U$'s biometric.
For instance,  $p_i \in \BT_U$ for a fingerprint represents the location and orientation of the fingerprint's \textit{minutiae}.
\textit{Minutiae} are regions in the fingerprint image where fingerprint lines start, end, merge and/or split.
In turn, each \textit{minutiae} $p_i$ is represented as:
\begin{equation}
p_i = (x_i, y_i, \theta_i)\,
\end{equation}
where $x_i$ and $y_i$ are Cartesian coordinates for the minutiae location in the fingerprint image and $\theta_i$ is the angle of 
minutiae orientation. In this paper, we focus on the fingerprint biometric modality, since fingerprint sensors are commonly 
found on commodity devices, such as laptops and smartphones. Nevertheless, similar encoding techniques are applicable to 
other biometric templates, such as iris scans~\cite{snuse_journal}.

\subsection{Fuzzy Extractors \& The Fuzzy Vault Scheme \label{sec:BG_FV}}
A Fuzzy Extractor~\cite{dodis2004fuzzy} (\FE) is a cryptographic primitive commonly used in biometric systems.
\FE can successfully extract the same randomness from different noisy samples of the same biometric as long as 
these samples are within a certain distance threshold. This fuzziness in the matching allows, for instance, to match biometric samples acquired using different sensors.
One popular \FE instantiation is the Fuzzy Vault scheme 
(\FV)~\cite{JS06} which is designed to work with \BTs represented by data point sets in Eq.~\ref{bt_format}.
An \FV scheme consists of two algorithms: $\FV_{GEN}$ and $\FV_{OPEN}$.
Given a biometric template $\BT_U$ the first algorithm generates the corresponding helper data \HD which hides a secret $k$.
Given another biometric template $\BT_U'$ and \HD, the second algorithm can successfully recover $k$ from \HD provided 
that $\BT_U'$ matches $\BT_U$.  The notion of \FV is captured in Definition~\ref{def:FV}.
Security of \FV relies on the infeasibility of the polynomial reconstruction problem~\cite{kiayias2002cryptographic}.
Definitions~\ref{def:completeness} and~\ref{def:security} formulate $\FV$'s completeness and (information theoretic) security.
\begin{figure}[ht]
\begin{mdframed}
\small
\begin{Definition}[\FV]\label{def:FV}
A Fuzzy Vault is defined as $\FV = (\FV_{GEN}, \FV_{OPEN}, \Phi)$,
where $\Phi$ is a set of parameters $\Phi = (d, GF(2^\tau), \texttt{MS}, \texttt{dist}, \texttt{w})$:\\
- $d$ is the polynomial degree;\\
- $GF(2^\tau)$ is a Galois Field of size $2^\tau$;\\
- $\texttt{MS}$ is a metric space;\\
- $\texttt{dist}$ is distance function defined over $\texttt{MS}$;\\
- $\texttt{w}$ is distance threshold;\\
$\FV_{GEN}$ and $\FV_{OPEN}$ are algorithms defined as follows:
\begin{compactitem}
\item $\FV_{GEN}$:
\begin{compactitem}
    \item \textbf{Inputs}: $k$ and $\BT_U$, s.t., $|k| = (d+1)\times\tau$.
    \item \textbf{Output}: $\HD$
\end{compactitem}
\item $\FV_{OPEN}$:
\begin{compactitem}
    \item \textbf{Inputs}: $\HD$ and $\BT_U'$
    \item \textbf{Output}: $k'$, s.t., $|k'| = (d+1)\times\tau$. 
\end{compactitem}
\end{compactitem}
\end{Definition}
\vspace{1mm}
\hrule
\vspace{1mm}

\begin{Definition}[\FV-Completeness]\label{def:completeness}~\\
$\FV = (\FV_{GEN}, \FV_{OPEN}, \Phi)$ is complete with $w$-fuzziness if for every possible $k \in GF(2^\tau)^{d+1}$ and every pair $\BT_U$, $\BT_U'$ with $\texttt{dist}(\BT_U,\BT_U') \leq w$: 
\begin{equation}
\FV_{OPEN}(\FV_{GEN}(k,\BT_U),\BT_U') = k
\end{equation}
with overwhelming probability.
\end{Definition}
\vspace{1mm}
\hrule
\vspace{1mm}

\begin{Definition}[\FV-Security]\label{def:security}~\\
$\FV = (\FV_{GEN}, \FV_{OPEN}, \Phi)$ is $p$-information theoretically secure if any computationally unbounded adversary with access to \HD is able to guess either, \BT or $k$, with success probability of at most $p$.
\end{Definition}
\end{mdframed}~
\end{figure}

$FV_{GEN}$ can be implemented by selecting a polynomial $P$ of degree $d$ defined over a field $GF(2^\tau)$ and encoding 
(or splitting) the secret $k$ into the $d+1$ coefficients ($a_i$) of $P$. The resulting polynomial is defined as:
\small
\begin{equation}
P_k(x) = \sum_{i=0}^{d}{a_ix^i}
\end{equation}
\normalsize
where coefficients $\{a_0, ..., a_d\}$ are generated from $k$ and can be used to reconstruct $k$. Since $P_k$ is defined over 
$GF(2^\tau)$, each coefficient can encode $\tau$ bits; this implies that size of a key that can be encoded is a function of the 
field size and the degree of the polynomial given by:
\begin{equation}
||k|| = (d+1) \times \tau
\end{equation}
After encoding $k$ as a polynomial $P_k$, each of the $M$ data points (features) in $\BT_U$ is evaluated in the polynomial 
$P_k$ generating a list of points in a two-dimensional space:
\begin{equation}
L_P = \{(p_1, P_k(p_1)), ..., (p_{M}, P_k(p_{M}))\}
\end{equation}
Note that the field must also be large enough to encode a single feature from $\BT_U$ as a single field element.
The resulting set $L_P$ is formed by only by points in the polynomial $P_k$.
In addition to $L_P$, a set of chaff points $L_S$ of size $N >> M$ is generated by randomly selecting pairs $(r_x,r_y) \sample GF(2^\tau)^2$, resulting in:
\begin{equation}
L_S = \{(r_{x,1},r_{y,1}), ..., (r_{x,N},r_{y,N})\}
\end{equation}
Finally, $L_P$ and $L_S$ are shuffled together using a random permutation $\pi_{\$}$ and the result is published as the helper data $\HD$:
\begin{equation}
\HD = \pi_{\$}(L_P + L_S)
\end{equation}
Note that $\HD$ also includes the set of public parameters $\Phi = \{F, d, l_P, H(k)\}$, where $F$ is the field over which $P_k(x)$ is 
defined and $d$ is its degree, $l_P$ is the size of $\BT_U$, i.e., the number of points in \HD that belong to $P_k(x)$, and $H(k)$ is a 
cryptographic hash of $k$ allowing one to verify if the correct secret is reconstructed using $FV_{OPEN}$\footnote{Using a hash function 
simplifies the implementation, but makes \FV's security computational in the size of the output of the hash. The scheme can be fully 
information theoretically secure by using error correcting codes.}.

The key idea behind security of the \FV scheme is that with $d+1$ distinct points $(p_i, P_k(p_i))$ (namely points on $P_k(x)$), one can 
interpolate $P_k(x)$, retrieve its coefficients and thus recover $k$. However, to find the \emph{right} $d+1$ points out of the $M+N$ points 
in the $\HD$ is very unlikely. With appropriate choice of $M$, $N$, and $d$ the success probability can be made negligible with 
respect to a desired security parameter.

To reconstruct $k$ from $\HD$ using a new biometric template $\BT_U'$, the $FV_{OPEN}$ algorithm applies a distance function 
(which must be defined according to the biometric type) to select $M$ points from \HD which have the shortest distance to the 
points in $\BT_U'$. If, out of the $M$ selected points, no less than $d+1$ points are indeed on the original the polynomial $P_k$, 
they can be used to interpolate $P_k$ and recover $k$. Otherwise, no interpolation with combinations of $d+1$ points out of $M$ 
correctly yields $P_k$ and therefore cannot recover $k$. To determine whether the resulting $k$ is correct, the algorithm compares 
its hash to $H(k)$ which is in the public help data.  $FV_{OPEN}$ rejects $\BT_U'$ if not equal; or accepts it otherwise.

The distance threshold $w$ can be used to tune the balance between the false acceptance rate (revealing $k$ to the wrong user) 
and the false rejection rate (refusing to reveal $k$ to the rightful user). \FV does not require ordered data points in the templates, 
and neither requires all data points to be in both sets. Only $d+1$ data points in $\BT_U'$ must be close enough to points in $\BT_U$.
The polynomial degree $d$ acts as an accuracy parameter allowing calibration of the scheme to reduce false acceptance by 
increasing the required number of matching data points.

In this work we use \FVs as a cryptographic building block to realize biometric-based \problem protocols.
As shown later in Section~\ref{sec:protocol}, \FV is used to cryptographically bind a random challenge 
chosen by \vrf to the biometric input in \protocol execution.

 \subsection{Hardware Architecture for Biometric Sensing with TEEs}\label{sec:hw}

 An advantage of biometric-based \problem is that in several types of modern devices, such as smart-phones and laptops, 
 biometric sensors exist and are directly connected (``hard-wired'') to the \rot exclusive memory itself, as depicted in Figure~\ref{fig:fp_tee}.
 it is usually the case that a biometric sensor (e.g., a fingerprint sensor) is directly hardwired to TEE exclusive memory.
 Therefore, the user's biometric input is not visible to untrusted (and potentially malicious) 
 software on that device, including the operating system. This means that an input biometric, cannot be obtained by an \sadv-controlled malware or OS on \devA, obviating the need for a trusted software path to be 
 verified by the \rot upon receiving the challenge. Nonetheless, our prototype implementation (see 
 Section~\ref{sec:implementation}) also considers the case where this hardware channel is not readily available. 
 In such a case, we show how to establish a secure channel between the biometric sensor and \rot with the help of a small trusted hypervisor.

\begin{figure}[!htbp]
\centering
  \fbox{\includegraphics[width=0.7\columnwidth]{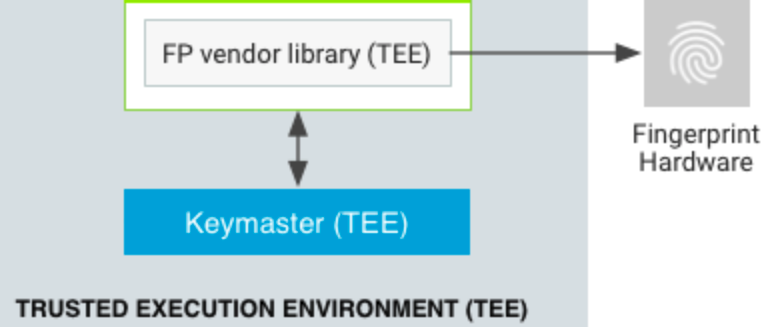}}
\centering \caption{TEE-Biometric hardware architecture of a typical Android device (adapted from~\cite{hw-fig})}\label{fig:fp_tee}.
\end{figure}
}

\section{Constructing an \problem Protocol} \label{sec:protocol}
\begin{figure*}[!ht]
\begin{center}
\fbox{
\scalebox{0.9}{
\procedure{}{%
\textbf{\vrf} \> \> \textbf{\prv}\\
\pcln  \BT_U \gets \mathsf{\BT.Sample}(U,\vrf) \> \> \rpki, \rski \gets Gen(\text{\prv}) \\
\pcln \chal \sample \{0,1\}^l\> \> \\
\pcln \HD \gets \mathsf{\FV_{GEN}}(\BT_U,\chal) \> \sendmessagerightx[3cm]{1}{\mathsf{\HD}} \> \\
\pcln \> \>  \BT_U' \gets \mathsf{\BT.Sample}_{}(U,\text{\devA})\\
\pcln \> \>  \chal' \gets \mathsf{\FV_{OPEN}}(\HD,\BT')\\
\pcln  \> \sendmessageleftx[3cm]{1}{\mathsf{\sigma, \rpki}} \>  \sigma \gets \mathsf{sign}_{\rski}(\chal')\\
\pcln \mathsf{verify}_{\rpki}(\sigma) \equiv \chal \> \> \\
}
}
}
\caption{\FV-based \problem protocol: \vrf decides whether \prv resides in \devA and, if so, learns its session public-key \rpki.}
\label{fig:FV_ROT_ID}
\end{center}
\end{figure*}
\begin{figure*}[ht]
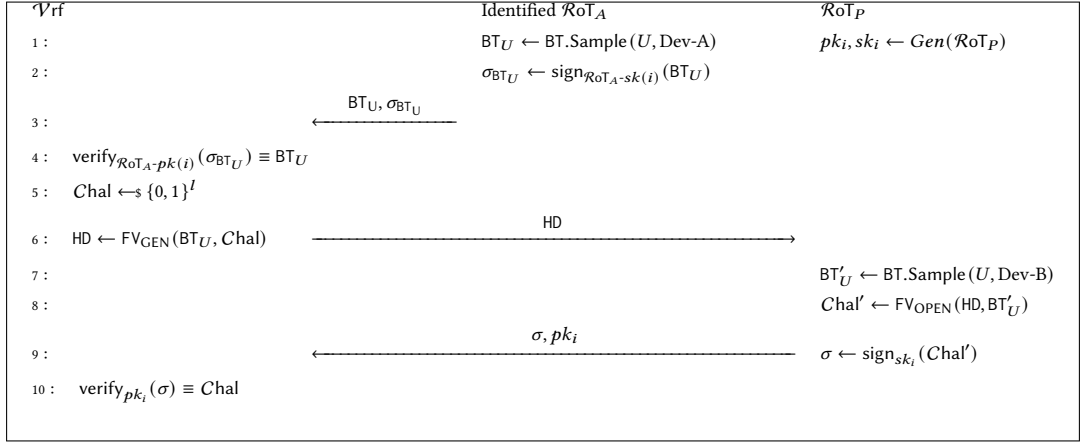

\begin{center}
\fbox{
\scalebox{0.9}{
\procedure{}{%
\textbf{\vrf}  \> \> \text{Identified \rotA} \> \> \textbf{\prv}\\
\pcln \> \> \BT_U \gets \mathsf{\BT.Sample}_{}(U,\text{\devA}) \> \> \rpki, \rski \gets Gen(\text{\prv})\\
\pcln \> \> \sigma_{\BT_U} \gets \mathsf{sign}_{\rskiA}(\BT_U)\> \> \\
\pcln \> \sendmessageleftx[2cm]{1}{\mathsf{\BT_U,\sigma_{\BT_U}}} \>  \> \>\\
\pcln \mathsf{verify}_{\rpkiA}(\sigma_{\BT_U}) \equiv \BT_U \> \> \\
\pcln \chal \sample \{0,1\}^l\> \> \\
\pcln \HD \gets \mathsf{\FV_{GEN}}(\BT_U,\chal) \> \sendmessagerightx[7cm]{3}{\mathsf{\HD}} \> \\
\pcln \> \> \> \> \BT_U' \gets \mathsf{\BT.Sample}_{}(U,\text{\devB})\\
\pcln \> \> \> \> \chal' \gets \mathsf{\FV_{OPEN}}(\HD,\BT_U')\\
\pcln  \> \sendmessageleftx[7cm]{3}{\mathsf{\sigma, \rpki}} \>  \sigma \gets \mathsf{sign}_{\rski}(\chal')\\
\pcln \mathsf{verify}_{\rpki}(\sigma) \equiv \chal \> \> \\
}
}
}
\caption{Proxy \problem protocol: \vrf is assisted by a previously identified \rotA (residing on \devA) to decide whether 
\prv resides on physical device \devB. \devA, \devB, and user $U$ must be physically co-located. \vrf can be remote.}
\label{fig:proxy_rot_id}
\end{center}
\end{figure*}
We now construct a biometric-based \problem protocol using \FVs and analyze its security.
We also present a Proxy \problem protocol that can be used to address \problem when \vrf is remote.

\noindent {\bf System Assumption:} We assume an authentic (not confidential!) channel between the biometric sensor and the RoT.  
In some types of devices (e.g., branded smartphones) similar channels are implemented in hardware in order to protect
the user's biometric data. Those channels are often claimed by vendors to be both confidential and authentic. Unfortunately, 
it has been recently shown that biometric data can still be leaked in clever ways\footnote{See: \url{https://www.blackhat.com/docs/us-15/materials/us-15-Zhang-Fingerprints-On-Mobile-Devices-Abusing-And-Leaking-wp.pdf}}, 
which means cuckoo attacks remain possible. 
In contrast, we believe that it is much harder to compromise authenticity of the channel, since the biometric sensor is hardwired to the RoT. 
Doing so would imply wholesale RoT compromise. Our scheme is dependent on channel authenticity and unclonability of fingerprints.  
For devices that do not have this kind of channel, we emulate it in software, by using a micro-hypervisor.

As discussed in Section~\ref{sec:RTI}, in a cuckoo attack on the challenge-response \problem protocol,
the adversary relays the challenge from \vrf. In a conventional challenge-response protocol, a correct response is 
formed based on two factors: the challenge and the prover's secret. Hence, to counter the challenge relay 
attack, we include the user in the loop as the third factor needed to produce a correct response.  
In particular, \vrf blinds the cryptographic challenge with the user's biometric by using an \FV scheme. 
\prv uses its biometric sensor to sample (presumably the same) biometric.
The user only provides her biometric to \devA's sensor, which can only be read by \rotA.
Therefore, the only \rot that can unblind the challenge is on \devA, which means \prv is \rotA.
Since the biometric given to both \vrf and \devA is the same, if \prv is not \rotA, \protocol for 
\devA fails. We now discuss the protocol in more detail.

\textit{\textbf{Remark:} We assume that protocol messages are exchanged over an encrypted and authenticated channel \vrf$\leftrightarrow$\prv. Note that this channel is established between \vrf and \prv, i.e., \vrf and {\bf some} \rot. Even though \prv has not been identified at this point, it is always possible to check whether \rpki was issued by some \rot. This is necessary to preserve confidentiality of \HD if a non-reusable \FE is used to implement the \problem protocol. (See Section~\ref{sec:limitations} for further discussion on \FE reusability.) A secure channel to some (trusted) \rot suffices to preserve confidentiality.}

\subsection{FV-based \problem}\label{sec:FV_protocol} 
Figure~\ref{fig:FV_ROT_ID} presents the \problem protocol based on the \FV scheme described in Section~\ref{sec:BG_FV}.
It assumes that \vrf and \devA are physically accessible to $U$.  $U$ participates in the protocol by providing the same 
biometric to the sensors of \vrf and \devA.

The protocol starts with \prv issuing an asymmetric key-pair and with \vrf sampling $U$'s biometric, thus resulting in the template 
$\BT_U$ (line 1). \vrf then generates a random $l$-bit challenge \chal, where $l$ is the security parameter (line 2).
Next, \vrf uses the \FV generation algorithm to obtain \HD where $\BT_U$ is the biometric and \chal is the secret.
\vrf sends \HD to \prv (line 3). $U$ also provides the same biometric to \devA. As a result, \rotA obtains $\BT_U'$ -- 
a new sample of the same biometric (line 4).

Note that the step in line 4 is crucial. Under the assumption of a secure channel between the 
fingerprint sensor in \devA and \rotA, $\BT_U'$ can only be obtained by the \rot residing in that device, i.e., \rotA.
If \prv does not reside in \devA, \sadv has to provide another biometric to \prv, i.e., from an accomplice person.
In such a case, due to \FV security, the reconstruction would result in an incorrect $\chal' \neq \chal$ with overwhelming 
probability $1-negl(l)$, for appropriate choice of \FV parameters as a function of $l$. Hence, it would not pass \vrf's signature verification (line 7). 
If verification succeeds, \vrf becomes convinced that \rpki is indeed issued by \rotA and \prv $\equiv$ \rotA.

Unlike \PA schemes, security of our biometric-based \problem scheme is based on \sadv's inability to forge $\BT_U$ 
and mount a successful cuckoo attack. Although \sadv controls entire software state of \devA (except for \rotA itself) and can access any memory 
outside of that reserved by \rotA, it cannot obtain $\BT_U'$ due to the secure channel between the fingerprint sensor and \rotA.

\noindent\textit{\textbf{Fingerprint Forgery:}} Fingerprints have been used as a biometric for a very long time and remain the most common
means of biometric authentication. There have been numerous successful attacks that surreptitiously obtain a user's 
fingerprints and then come up with various contraptions to fool fingerprint sensors. Clearly, the proposed protocol and its variations
will fail if the biometric template used in a \problem protocol execution is stolen and reproduced before hand. 
However, the protocol does not require a pre-determined fingerprint or the user. Hence, the fingerprint forgery attack 
may not always succeed.

\begin{figure}[ht]
\begin{mdframed}
\small
\begin{Theorem}\label{th:fv_roti_completeness}
\FV-based \problem protocol (Figure~\ref{fig:FV_ROT_ID}) is complete according to Definition~\ref{def:RTI_comp} 
as long as \FV is complete according to Definition~\ref{def:completeness}.
\vspace{1mm}
\begin{Proof}
In an honest execution of the protocol \prv resides in \devA, i.e.: $\text{\rpki} \gets Gen(\text{\rotA})$.\\
Since, \prv resides in \devA, \vrf and \prv (i.e., \rotA) receive $\BT_U$ and $\BT_U'$ such that $dist(\BT_U, \BT_U') \leq w$.
It follows from Definition~\ref{def:completeness} that:
\begin {equation}
\begin{split}
 & \HD \gets \FV_{GEN}(\BT_U, \chal) \rightarrow \\
 & Pr[\FV_{OPEN}(\HD,\BT_U') = \chal] > 1-negl(l) \rightarrow \\
 & Pr[\sigma \equiv sign_{\rski}(\chal)] > 1-negl(l) \rightarrow \\
 & Pr[verify_{\rski}(\sigma) \equiv \chal = 1] > 1-negl(l)
\end{split}
\end {equation}
\end{Proof}
\end{Theorem}
%
\vspace{1mm}
\hrule
\vspace{1mm}
\small
\begin{Theorem}\label{th:fv_roti_security}
\FV-based \problem protocol (Figure~\ref{fig:FV_ROT_ID}) is secure according to Definition~\ref{def:RTI_sec}, as long as
\FV is $p$-information theoretically secure as in Definition~\ref{def:security} and \FV parameters are chosen such that and $p = negl(l)$.
\vspace{0.3mm}
\begin{Proof}
In this case, \prv does not reside in \devA i.e.: $\neg(\text{\rpki} \gets Gen(\text{\rotA}))$.\\
Therefore, it must be the case that \vrf and \prv receive $\BT_U$ and $\BT_U'$ such that $dist(\BT_U, \BT_U') > w$.
Assuming that \sadv is unable to forge $sign_{\rski}(.)$ with more than $negl(l)$ advantage, it follows from Definition~\ref{def:security} that:
\begin {equation}
\begin{split}
 & \HD \gets \FV_{GEN}(\BT_U, \chal) \rightarrow \\
 & Pr[\FV_{OPEN}(\HD,\BT_U') = \chal] = p = negl(l) \rightarrow \\
 & Pr[\sigma \equiv sign_{\rski}(\chal)] = negl(l) \rightarrow \\
 & Pr[verify_{\rski}(\sigma) \equiv \chal = 1] = negl(l)
\end{split}
\end {equation}
\end{Proof}
\end{Theorem}
\end{mdframed}
\end{figure}

As  mentioned above, security of the protocol in Figure~\ref{fig:FV_ROT_ID} depends on that of the \FV scheme.
Completeness and security of this protocol are stated in Theorems~\ref{th:fv_roti_completeness} and~\ref{th:fv_roti_security}, 
respectively. In both completeness and security arguments, we assume that whenever two samples are taken from the 
same biometric they are within a certain distance threshold. Conversely, we assume that two samples of different biometrics 
are beyond that threshold. In other words, $dist(\BT_U, \BT_U') \leq w\; \iff \; \BT_U$ and $\BT_U'$ are samples of the same biometric.
In practice, validity of this assumption depends on the accuracy of the biometric matching procedure, including the distance function 
$dist$, the distance threshold $w$ and the degree of \FV polynomial. Our choice of parameters are based on previous work 
 on these issues and are discussed in Section~\ref{sec:implementation}.
Accuracy results obtained with such parameters are discussed in Section~\ref{sec:limitations}.

\subsection{Proxy \problem Protocol}
%
The protocol in Section~\ref{sec:FV_protocol} requires \vrf, and \devA to be physically accessible to $U$, since $U$ must
provide her biometric sample to both \vrf and \devA.  To cope with scenarios where \vrf (e.g., a server) is not easily approachable, we suggest to use a proxy \devA with its \rotA previously identified, in order to assist \vrf in identifying 
\rotB.

Suppose that $U$ now carries \devA to the location of \devB.
Figure~\ref{fig:proxy_rot_id} shows a protocol for using \devA to assist \vrf in remotely identifying \rotB of \devB. 
The main idea is for  \rotA to act as an interface of \vrf. It captures $U$'s biometric and forward it to \vrf via an authenticated and secret network channel. 
The same biometric is also used as a challenge to \devB, which runs the rest of the protocol with \vrf.

The security of the \FV-based \problem protocol in Section~\ref{sec:FV_protocol} implies the security of the 
proxy \problem protocol. We note that \devA is \emph{not} a trusted device as used in \cite{DPKC18}. 
Its software, including the OS, could be compromised, while its \rotA is trusted, which is consistent with 
the basic protocol. Hence, both protocols provide the same level of security.  


As discussed earlier, lack of \problem violates the assumption that \rot resides on the physical device of interest, thus 
undermining security of any application dependent on that assumption. The Proxy \problem is itself a good example of 
such an application. It relies on the assumption that biometric sampling is performed on \devA -- the device in possession of 
authorized user $U$. Therefore, identification of \rotA is crucial to overall security of this application.

\section{Prototype \& Evaluation}
\label{sec:implementation}
\subsection{\BT Extraction \& \FV Parameters}
%
%
\BT extraction generates a biometric template from a fingerprint image.
As discussed in Section~\ref{sec:preliminaries}, each data point $p_i \in \BT$ is the position and 
orientation $(x_i,y_i,\theta)$ of a fingerprint minutiae. To extract the \BT we use NIST Biometric Image Software (NBIS)~\cite{nbis}.
NBIS returns a set of identified minutiae points with corresponding confidence levels.
From NBIS output, we select 20 points with the highest confidence and encode them as data points in $GF(2^{24})$.
In our prototype, \FV's \HD is composed of 20 fingerprint data points mixed with 200 random chaff points.
The \FV polynomial degree is set to 9. Finite field operations are implemented using the Number Theory Library (NTL)~\cite{ntl}.

\begin{figure*}[!hbtp]
\centering
  \subfigure[Fingerprint pre-processing.]
  {\includegraphics[height=1.7in,width=0.32\textwidth]{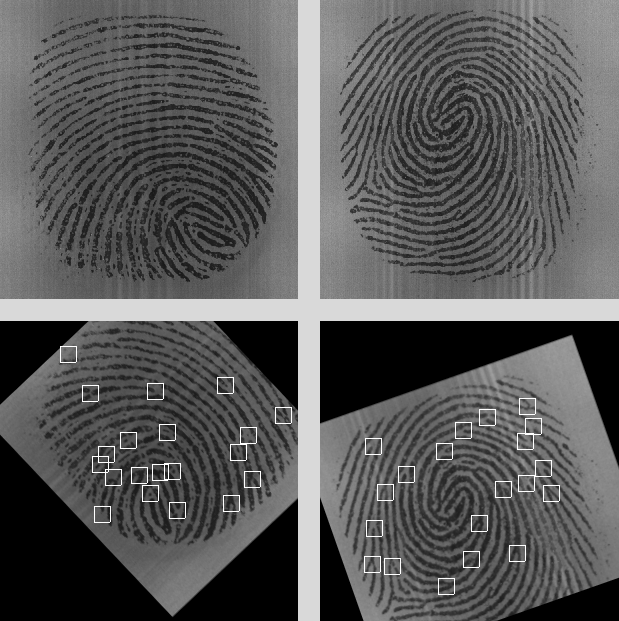}\label{fig:fp_extraction}}
  \subfigure[Hardware Setting]
  {\includegraphics[height=1.7in,width=0.32\textwidth]{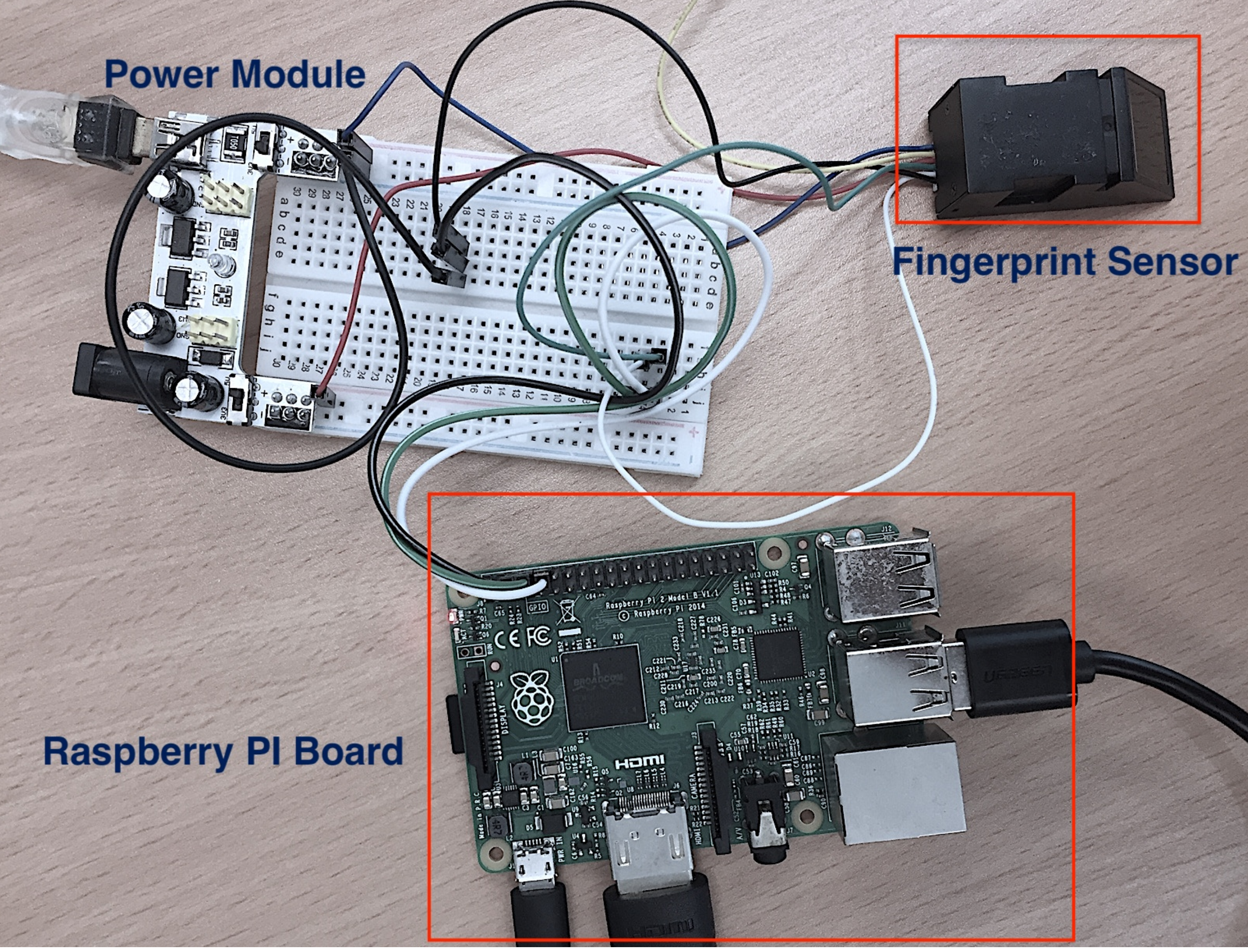}\label{fig:setup}}
  \subfigure[Hypervisor Based \rot. Arrows illustrate execution flow; shaded area denotes untrusted software.]
  {\includegraphics[height=1.7in,width=0.32\textwidth]{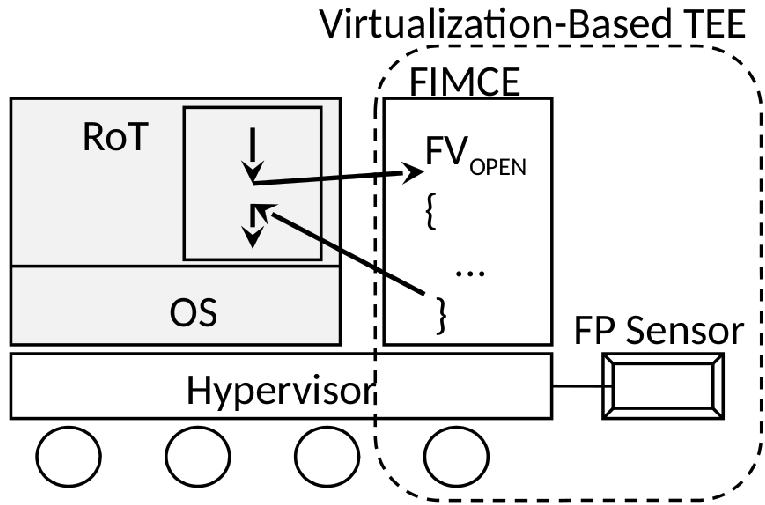}\label{fig:tee}}
\caption{Hardware and software components of our prototype.}
\end{figure*}

In $FV_{OPEN}$, the candidate minutiae points are selected from the \HD based on their distance to minutiae points in the 
new template $\BT'$ sampled from the user. Similar to~\cite{nandakumar2007fingerprint}, we use a distance function 
between $p_i\in\HD$ and $p'_j\in\BT'$ defined as:
\begin{equation}
D(p_i, p'_j) = \sqrt{(x_i-x'_j)^2 + (y_i-y'_j)^2} + \beta \times \Delta(\theta_i,\theta'_j)
\end{equation}
where $p_i=(x_i,y_i,\theta)$, $p'_j=(x'_i,y'_i,\theta')$, and $\Delta(\theta_i,\theta'_j) = min(|\theta_i - \theta'_j|,360 - |\theta_i - \theta'_j|)$.
Parameter $\beta$ controls the degree of importance given to minutiae orientation in computation, as compared to the euclidean distance 
between the points. A data point $p_i$ is selected if $D(p_i,p'_j) < w$ for some point in $p'_j\in\BT'$.
As described in~\cite{nandakumar2007fingerprint}, parameters $\beta$ and $w$ must be empirically calibrated to 
yield the best accuracy results. Our parameters are empirically calibrated to: $\beta = 0.2$ and $w = 20$.
To improve accuracy results for noisy fingerprint readings before extracting the template, during the biometric sampling, 
we run the fingerprint pre-alignment algorithm from~\cite{tarp}. Figure~\ref{fig:fp_extraction} illustrates the result of the 
template extraction for two pre-aligned fingerprint images. White squares highlight the minutiae points detected in these fingerprints.
We discuss the accuracy of this implementation in Section~\ref{sec:accuracy}.

\textit{\textbf{Remark:} We implement our own \BT extraction to have a fully working prototype 
and report on its accuracy. We stress that accuracy of the underlying \BT extraction technique is orthogonal and not affected by 
the \problem setting considered in this work.}

\subsection{Prototype} 
Due to the close environment of hardware-based TEEs with fingerprint sensing (commonly found on mobile phones), we implement the prototype of FV-based \problem on a development board connected with an external fingerprint sensor.
The sensor collects user fingerprints and also provides an interface to export the data to a secure storage inaccessible to applications and the operating system. We build a hypervisor-based secure execution environment (software-based \rot) to run \prv steps in the FV-based \problem protocol. 
\subsubsection{Hardware Setting}
Figure~\ref{fig:setup} shows the hardware setting of our prototype. An FMP12 Optical Fingerprint sensor is connected to the Raspberry Pi 2 development board with four Cortex-A7 CPU cores at 800 MHz and 1 GB main memory. It runs Debian Linux with kernel version 3.18.8. Software on the board can use a serial port mapped at physical address 0x3F201000 to issue commands to the fingerprint reader and read the collected data.  
\subsubsection{Virtualization Based \rot}
We harness virtualization techniques to build an \rot secure against attacks from the operating system. Our secure environment shown in Figure~\ref{fig:tee} is implemented by following the approach proposed in \cite{fimce} which designs a fully isolated minimal computing environment (FIMCE) on a multicore x86 platform. We develop a bare-metal ARM hypervisor running in the processor's Exception Level 2 (EL2) which is more privileged than the levels for the OS and applications. After launched on the Raspberry Pi board, the hypervisor configures the permission bits in the Stage-II translation table to block the OS and applications from accessing the serial port used by the fingerprint sensor. Hence, the adversary cannot access the fingerprint sensor to issue commands or steal fingerprint images. When available, a secure boot module can be used to assure that this configuration is properly set at boot time.

Upon receiving a request, the hypervisor creates a fully isolated computing environment consisting of a CPU core and a reserved physical memory region for the sensitive function to run. The CPU configuration ensures that maskable interrupts are not delivered the core and non-maskable interrupts (NMIs) are trapped to the hypervisor. Thus, the untrusted OS cannot tamper with the environment via memory accesses or interrupts. 
 
Running in the isolated environment is the code implementing \prv logic in the \problem protocol.  Its signing key $sk$ is stored in the hypervisor memory.
To generate the response, it requests the hypervisor to run $\mathsf{FV_{open}}$ and $\mathsf{sign_{sk}}$ in the FIMCE environment at runtime. The code of these two functions are self-contained without issuing system calls so that the executions do not depend on any untrusted code and data outside of the isolated environment. Considering that these two functions are for memory-resident computations without involving I/O operations, system calls are avoided by statically allocating the needed memory buffers. Note that an ARM CPU does not allow a user privilege code to issue hypercalls. Hence, we retrofit the OS with a special system call handler which issues the hypercall on behalf of \prv. 
%
%

\subsubsection{Evaluation}
Code complexity is shown in Table \ref{tab:code}.
We measured CPU execution time for $\mathsf{FV_{OPEN}}$ and $\mathsf{sign_{sk}}$ 
within the virtualization-based \rot and normal user space on the Raspberry PI board. 
Results are reported in Table~\ref{tab:time} below. 
\begin{table}[!hbtp]
\centering
\scalebox{0.90}{
\begin{tabular}{|l|c|} \hline \cline{1-2}
{\bf System Component} & {\bf LoC} \\ \hline \cline{1-2}
ARM hypervisor & $456$ (C) and $906$  (Assembly)  \\ \hline
Self-contained $\mathsf{FV_{OPEN}}$ &  $701$ \\ \hline
Crypto library (incl. RSA and hash functions) & $5,032$ \\ \hline 
\end{tabular}
}
\vspace{1mm}
\caption{Code Complexity (in LoC).}
\label{tab:code}
\end{table}%
\begin{table}[!hbtp]
\centering
\scalebox{0.990}{
\begin{tabular}{|c|cc|} \cline{2-3}
 \multicolumn{1}{c|}{}  & $\mathsf{FV_{OPEN}}$ & $\mathsf{sign_{sk}}$ \\ \hline
 Native environment & 848.7 & 79.2 \\ \hline
 Virtualization-based \rot & 1143.51 & 75.6 \\ \hline 
\end{tabular}
}
\vspace{1mm}
\caption{CPU time comparison. Average of out of 1000 executions (time in {\em ms}). 
Variance was negligible and omitted} 
\label{tab:time}
\end{table}
We note that time differences are not large. In fact, the RSA signing operation has a slight performance 
advantage when running in the \rot. The reason might be its exclusive use of the CPU core since interrupts are blocked. 

\section{Practical Considerations}\label{sec:limitations}
We now discuss some practical issues relevant to the proposed \problem protocol.
\subsection{Biometric Sensor Availability}
One limitation of our general approach is the requirement for a biometric sensor hardwired to the \rot. Our prototype shows how 
this requirement can be circumvented -- the protocol can be securely deployed on devices not equipped with 
embedded biometric sensors by using a stand-alone biometric sensor and a trusted micro-hypervisor to 
emulate a hardware direct channel between the sensor and \rotA.

Nonetheless, we recognize that it might be beneficial  to remove this hardware dependence.
In particular, it would be interesting to develop new \problem protocols that use other types of 
physical challenges through other sensors that (similar to biometrics) are hard to clone/replay.
In particular, developing alternative \problem based on other sensors that might be available on commodity 
devices and evaluating their usability trade-offs is an interesting future direction.

\subsection{Biometric Confidentiality}
One concern with the proposed protocol is confidentiality of the biometric data used in the protocol. 
%
Even though \devA might be compromised, 
the biometric sample is read directly by trusted \rotA. In other words, confidentiality of the user's 
biometric vis-a-vis \devA is guaranteed, assuming that \rot hardware tamper-resistance is preserved.
The same applies to \vrf, if it is also equipped with a \rot. Otherwise, the owner of \vrf should be the same
as the user providing providing the biometric.

\subsection{Fuzzy Extractor Issues}
Statistical and reusability attacks are well-known issues of several \FE constructions, including fuzzy 
vaults used in our prototype. The former is the biometric analog to dictionary attacks on passwords.
It analyses the distribution of minutiae in human biometrics and uses this information to extract 
\BT or \chal from \HD. The latter applies to non-reusable \FEs. In such cases, obtaining two instances 
$\HD_1$ and $\HD_2$, generated from the same biometric allows reconstruction of \BT in clear.

We note that these attacks are a serious concern for \FE-based biometric authentication where 
\HD appears in clear. Whereas, in our case, the problem is obviated by transmitting \HD over a secure channel to \prv.
In particular, we do not use \FEs for biometric confidentiality (since they are not necessary to achieve that purpose).
They are used such that \vrf can always embed a fresh challenge \chal into the ``biometric-based'' challenge, 
preventing replays of previous \problem executions with the same biometric on other \rot, e.g., \rotAdv.

\subsection{Usability}
As mentioned earlier, usability is a problem with sight-based presence attestation, along with its reliance on 
precise timing. Recall that location- and scene-based presence attestation schemes incur lower user burden.
However, they also offer much lower security. Meanwhile, user burden in our protocol amounts to performing two 
biometric samplings: one with \vrf and one with \devA. (Moreover, the user can pre-enroll his fingerprints with \vrf
well ahead of time.)This type of user interaction is common for authentication 
purposes and typically considered more convenient than other authentication means, such as entering a PIN or 
password. Therefore, we consider usability of biometric-based \problem protocol to be quite reasonable.

\subsection{Accuracy}\label{sec:accuracy}
\begin{figure}[!htbp]
\centering
  \includegraphics[height = 2.0in, width=0.8\columnwidth]{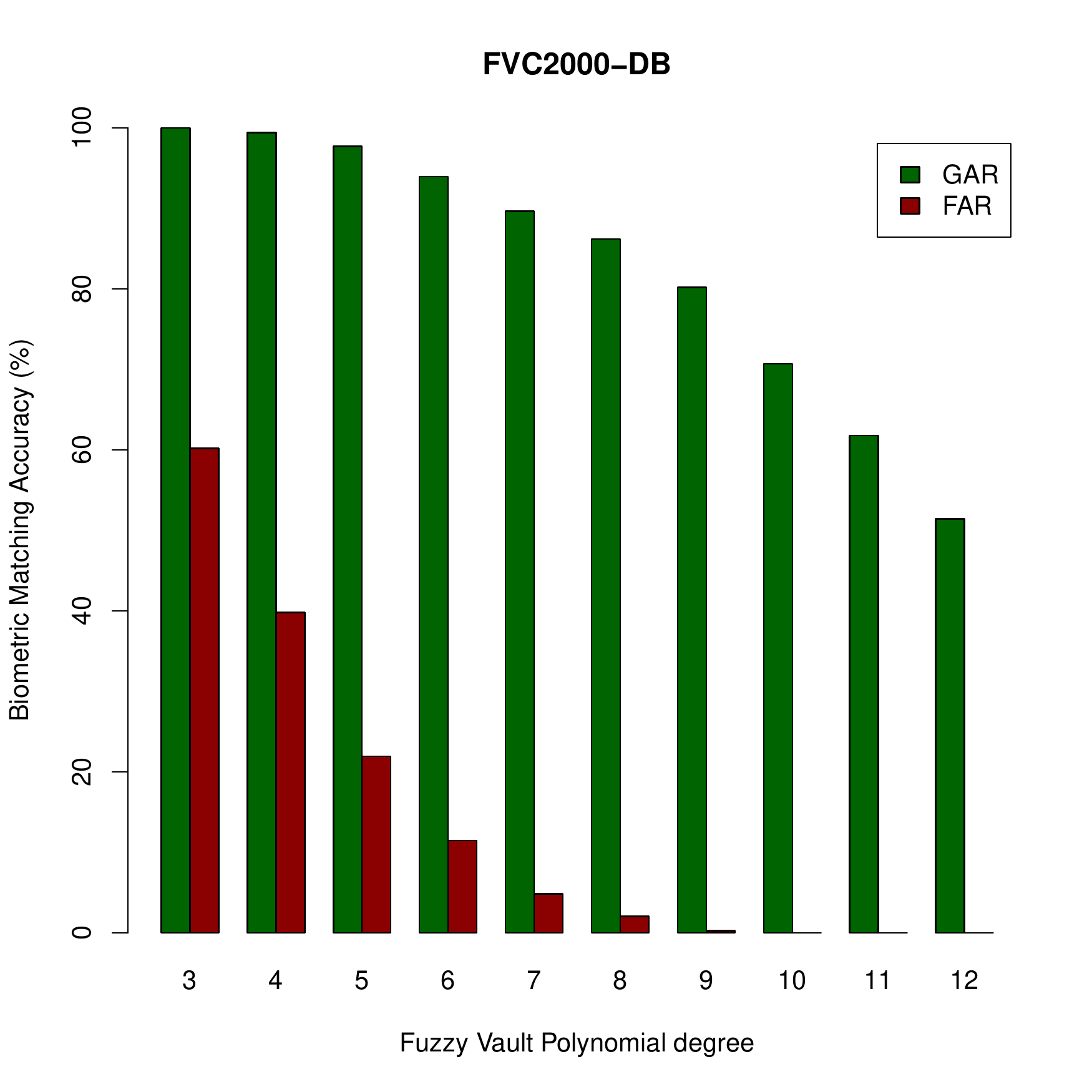}
\centering \caption{Accuracy of biometric matching in our prototype}\label{fig:accuracy}.
\end{figure}

Accuracy of the underlying biometric matching is not affected by our use-case.  Improving its accuracy is 
an orthogonal effort. Nonetheless, for completeness, we report on the accuracy considering the implementation 
used in our prototype. Similar accuracy analysis for biometric matching using fuzzy vaults (also considering other 
biometrics modalities) can be found in~\cite{nandakumar2007fingerprint,snuse_journal,iris_fv}.
We report on our prototype's accuracy considering metrics for:\\
-- \textbf{Genuine Acceptance Rate (GAR):} Percentage of biometric samples correctly matched to other samples 
acquired from the same biometric.\\
-- \textbf{False Acceptance Rate (FAR):} Percentage of biometric samples incorrectly matched to any sample 
not acquired from the same biometric.\\
We conducted accuracy experiments using FVC2000 publicly available fingerprint database (database and further information 
available at: \url{http://bias.csr.unibo.it/fvc2000/}). FVC2000 includes multiple fingerprint 
images (10 different noisy images of each fingerprint) acquired using 4 types of low-cost biometric sensors.
As discussed in Section~\ref{sec:BG_FV}, the \FV polynomial degree allows configuring the number of matching 
data points in two biometric samples necessary to consider that the samples belong to the same user.
Therefore, accuracy results are presented as a function of \FV polynomial degree in Figure~\ref{fig:accuracy}.
According to the results in Figure~\ref{fig:accuracy}, for a security-critical task such as \problem, an ideal choice would 
be degree $9$ with nearly zero false acceptances. The same degree results in GAR of 80\%, meaning that 1 out of 5 times a 
genuine \problem execution would fail and the user would need to try one more time.
%

\section{Related Work}\label{sec:rw}
In this section we summarize topics related to \problem, except  \PA~\cite{presence_att} which was already 
discussed in Section~\ref{sec:pres_att}.

\textbf{\textit{Cuckoo Attacks}} were thoroughly introduced and formally modeled in~\cite{parno2011bootstrapping}.
Several potential solutions were analyzed under that model and, among them, secure hardware channels between \devA I/O interfaces the \rotA were considered as the preferred method.
As discussed in Section~\ref{sec:RTI}, even direct channels can be circumvented by a Cuckoo \sadv that deploys its own \emph{accomplice challenger} to replay \vrf messages through the appropriate channel.
To tackle this problem, our biometric-based approach explores the uniqueness of biometrics as a physical 
unclonable challenge, in addition to the existing secure channel between the biometric sensor and the \rot.

\textbf{\textit{Distance Bounding (DB)}} is a promising approach for addressing the
\problem problem. With recent advances~\cite{leu2019message,singhuwb,235453}, DB could allow \vrf to precisely establish maximum distance (bound) to the untrusted \prv.
Basically, if each device is equipped with DB facilities (a special radio and a high-precision clock)
and \prv has a secure hardware channel to DB in its housing device, then the user can simply make sure that no other device is within the reported bound, e.g., 20-30 cm.
However, several obstacles (discussed in Section~\ref{sec:RTI}) must be overcome before DB can be used for \problem.
 
\textbf{\textit{User Trust Bootstrapping}} allows the user to establish trust on her device.
TrustICE~\cite{sun2015trustice} uses a hardware approach and uses an LED under exclusive control of \rot.
The light signal emitted by this LED is used to convince the user that the device has an active \rot.
Other approaches~\cite{danisevskis2015graphical,lange2013crossover} reserve a fraction of \dev 
screen to communicate the state of the trusted component to the user.
While these approaches succeed to communicate the state of \rot in a given device, they do not provide 
identification of corresponding public keys.
 
\textbf{\textit{Device Pairing}} is the problem of initializing a secure (usually wireless) channel between 
two previously unfamiliar devices, without any trusted third party. Many device pairing protocols have been
proposed, relying on various physical properties~\cite{saxena2006secure,soriente2008hapadep,jung2011device}. 
The main difference between \problem and device pairing is that, in the former,  one of the devices (\devA) is potentially 
compromised and is therefore subject to cuckoo attacks. In contrast, device pairing mainly considers evil twin attacks.

\textbf{\textit{Remote Attestation}} is an \rot-enabled security service that allows \vrf to measure software state of 
applications running on \dev. In recent years, several remote attestation techniques and 
architectures~\cite{haldar2004semantic,barbosa2016foundations,smart,hydra,vrased,simple} were proposed, targeting different 
platforms and offering 
different types of guarantee. While remote attestation enables malware detection on a remote \dev, 
it cannot be used as a means to solve \problem by ensuring that \dev is in a malware-free state. This is because 
remote attestation itself requires mitigating the \problem problem, i.e., making sure that a remote attestation protocol 
indeed executes on \dev before it can be used to ensure that \dev is malware-free.

\textbf{\textit{Biometrics}} are widely used in user authentication~\cite{snelick2005large,FIDO,burger2001biometric,bhagavatula2015biometric} 
and identification~\cite{haghighat2015cloudid,yuan2013efficient} systems. Fuzzy extractors are typically deployed to preserve 
biometric template confidentiality in the back-end of these systems~\cite{snuse}. To the best of our knowledge, this paper is the first proposal 
to use biometrics and fuzzy extractors to convey an unclonable challenge and assist in the identification of an \rot. 
\vspace{-0.8mm}
\section{Conclusion}\label{sec:conclusion}
This paper introduced and analyzed the \problem problem, which occurs whenever an \rot is used to implement a 
security service that depends on physical IO devices (sensors and actuators) and relies on the assumption of
\rot residing in a specific physical device. To address this problem we proposed an \problem protocol based on the 
difficulty of cloning biometrics in real time.
It uses the biometric as a challenge in the \problem protocol and 
relies on the existence of a hardware channel between biometric sensors and TEEs -- a feature
already available on some current devices.
We also demonstrated a prototype implementation of our 
approach.


\balance
\bibliographystyle{ieeetr}
\bibliography{references}

\end{document}